# First and second order electromagnetic equivalency of inertial systems, based on the wavelength and the period as speed-dependant units of length and time

G. Sardin
gsardin@ub.edu

**Abstract**

The cause for first and second order electromagnetic equivalency of inertial systems is approached from a different point of view than that of special relativity. While special relativity applies dilatation to time and contraction to space itself, *the proposed framework applies restrictively these effects to the units of space and time, embodied in the beam wavelength and period, as perceived on the inertial system due to the Doppler effect*. It is not space and time themselves that would actually vary but the *electromagnetic units* of space and time constituted by the wavelength and period. Mathematical constructs accept indifferently both setting out, which provide identical results, so their valuation must be made on physical grounds. The reader is invited to ponder on interpreting physical reality through the contraction of space and dilatation of time or through the variation of the *natural units of space and time provided by the wavelength and the period*. These two issues, mathematically equivalent, lead however to drastically different conceptions of the physical world.

*Introduction*

Electromagnetic equivalency of inertial systems, as inferred from optical experiments such as Michelson's archetypal one and its new versions relying on optical cavities, is addressed from a new perspective of generalised application to the whole electromagnetic framework, that accounts for the Doppler effect. An extension of this perspective has been applied to the dependance on speed of the rate of atomic clocks and of the life-time of elementary particles such as the muon and the pion, seen as elemental clocks (*). First-order equivalency, corresponding to the invariance from speed of the light spots position on the beam-splitter, most often overlooked presumably being given for granted, is here carefully analysed in view of its crucial relevance. Second order equivalency, i.e. interferometric invariance, is interpreted in terms of units of time and space provided by the period and the wavelength, and their dependence on speed, instead of inferring time dilatation and contraction of space itself. Time dilatation and space contraction are viewed as mere mathematical wits, void of any actual physical reality. Let us underline that it is indeed highly incorrect to extract them from their mathematical context and to project them as factual physical features. In effect, the interferometric optical experiments do not actually measure any time dilatation or space contraction, but rely instead on the properties of light beams in regard to the conveying system motion. These set-up do not directly measure time but provide only a differential measure of time, i.e. the two beams time-of-flight difference through the phase. Since what is observed is the invariance of the phase, this null result should be accurately and restrictively interpreted in terms of electromagnetic properties, instead of those of time and space.

Special relativity has interpreted the equivalency of inertial systems appealing to space contraction and time dilatation [1-5]. Here it will be demonstrated that it is not necessary to resort to these effects to give account of their equivalency. But let us make previously a very brief introductory review. The recurring experimental failure to detect effects due to the earth motion led to controversy between theorists such as G.P. Fitzgeralt, H.A. Lorentz and J.H. Poincaré, until 1905 when A. Einstein proposed a special mathematical construct, stating that the Galilean transformation should be supplanted by a new space-time transformation, called the Lorentz transformation. In regard to experimental back-up it is not overt that Einstein was straightly impelled in his setting out by the Michelson experiment, but in his 1905 publication he alludes to "*the unsuccessful attempts to discover any motion of the earth relative to the light medium*".

In fact Maxwell's theory of electrodynamics does not satisfy the principle of Galilean relativity. Maxwell equations foresee that the speed of light through free space is a universal constant c, but this will not be fulfilled in any inertial system defined by the Galilean transformation. Maxwell thought that the electromagnetic waves were carried by a medium, so that his equations would only hold in a restricted group of Galilean inertial frames, i.e. in those at rest with respect to the referential spatial medium. The theory of relativity does not actually solve the antinomy between absolute and relative space, it just discards the former. On experimental grounds, even though the earth has a velocity of 30 km/s relative to the sun and about 230 km/s relative to the centre of our galaxy, all attempts to measure the earth velocity with respect to the space medium failed. Nevertheless, using instead the electromagnetic spectrum of the cosmic microwave background, latterly discovered and which can suitably supplant for practical means the early hypothetical spatial medium as a universal extended reference, it has been unveiled that our galaxy has a drift velocity of about 600 km/s with respect to it, outcoming a net drift velocity of our solar system of about 370 km/s.

In order to study the significance of a given experimental test of special relativity, it is useful to employ a framework allowing its transgression. A popular one for this purpose is the "test theory" of Mansouri and Sexl [6]. In a similar way, in order to check the adequacy of the digressing framework let us use an approach consisting in establishing a conceptual layout transgressing that of special relativity. If the postulates of special relativity are presumed to be correct, transgressing ones should lead to results not fitting with experimental ones. In other words, since from the initial premise, right predictions are the patrimony of special relativity, transgressing fundamentals should hence lead to wrong predictions [7-12]. However, what happens if the surmised postulates, incompatible with those of special relativity, equally provide the same correct predictions? But let us first define the clauses of the transgressing framework and let us check if its predictions fit experimental results.

*The transgressing postulates are:*

(1) In all optical experiments, the referencial frame must be light itself. So, experimental results should be analyzed in terms of the relationship between light and the inertial observer, i.e. of the way the fundamental properties of light are perceived on the inertial system due to its motion relative to the beam.

(2) The *absolute* speed of light (c) is an intrinsic property, which is constant exclusively in relationship to space itself. This implies that c is independent of the source speed.

(3) The Doppler effect should always be accounted for since the proper inertial system is actually moving within space and has a velocity relative to that of light. Hence, the *relative* speed of light is not at all independent of the observer speed.

*Let us express two corollaries:*

(1) First order optical equivalency, as derived from Michelson-like experiments (fig.1), implies the tangential speed ($c_x$) of the reflected light beam and that of the inertial system to be equal ($c_x = v$).

(2) Second order optical equivalency, i.e. interferometric invariance (fig.1 and 2), implies that the wavelengths perceived by the inertial observer are tuned by the Doppler effect, in a way that the two split beams meet back with the same phase.

## 1. First order equivalency

In monographs and textbooks dealing with special relativity, reference to Michelson's experiment only addresses second order equivalency, i.e. the invariance of the interferometric pattern. First order equivalency, i.e. the invariant concurrence of the beam spots at the same steady position on the beam-splitter under any directional change, after the detached journey of the two split beams, is systematically overlooked (fig.1). However, this is not an obvious fact and it should be appraised carefully. In effect, once the two split beams have been adjusted such that they impinge back on the beam-splitter at the very same point this setting endures with time and under any rotation of the interferometer. Hence, it has been inferred that the position of the spots is independent of any change of the conveying inertial system motion.

This crucial and not at all trivial point deserves an acute analysis. In effect, after the incident beam impinges on the beam-splitter, the split parts ensue self-reliant round-trip trajectories, but both still coincide back at the very same initial position on the beam-splitter, despite it has drifted away during their time-of-flight, which moreover differ. Most particularly, the fact that the reflected beam, after departing from the beam-splitter, meets it back at the same point independently of the motion of the conveying inertial system, and thus of the resultant variable spatial dislodge of the beam-splitter, must necessarily be ruled by some very skilful physical law.

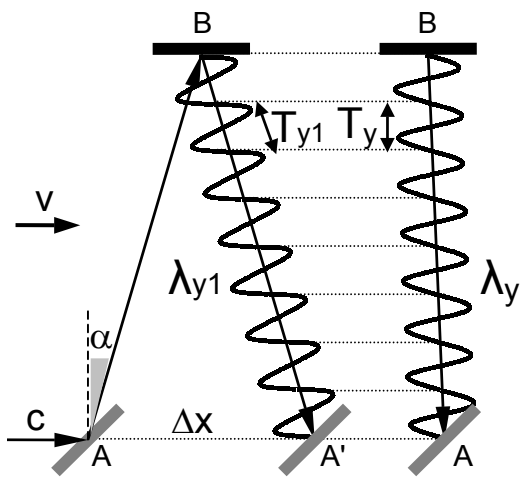

**Fig. 1:** Actual path, wavelength and period of the reflected beam (left), and as perceived by the inertial optical set-up (right)

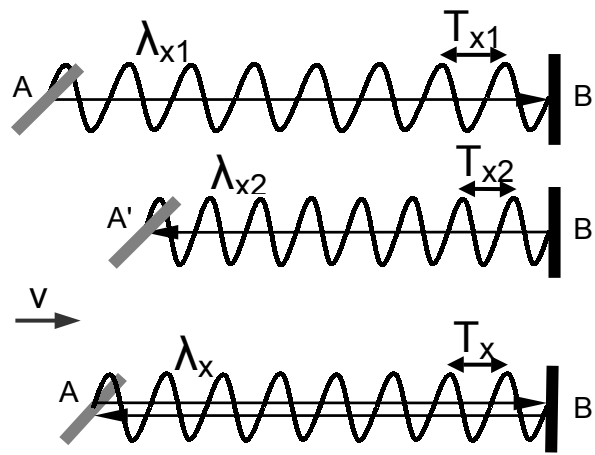

**Fig. 2:** Actual path, wavelength and period of the transmitted beam (upper two graphs) and as perceived by the inertial optical set-up (lower graph)

Let us stress that the law governing this invariance of the spots position, obeying a first order dependence on speed, appears to be ruled by the transfer of momentum, which in turn implies that the tangential speed of the reflected beam must be equal to that of the inertial system. Without this fulfilment the reflected beam would be unable to catch back the beam-splitter at the precise initial spot position. In other words, this requires the reflected beam to be slightly deflected in the direction of the system motion, so that its tangential speed ($c_x$) balances that of the system ($c_x = v$).

### *1.a. Equality of the drifts from the reflected beam and the beam-splitter*

First order equivalency necessarily implies that no effect related to the inertial system speed can ensue. Therefore, if the mirror on the Y axis is adjusted so that the beam reflects at exactly the same position on the beam-splitter, then any variation of the speed should leave unaltered the spots position. In other words, whatever the speed the reflected beam on the beam-splitter must impinge back on it, after its round-trip time-of-flight (t) at the very same place as at first. Since the beam-splitter has meanwhile drifted of a distance $\Delta x = v.t$, this implies both incident and reflected beam spots to be spacially separated by the same distance $\Delta x$ (fig.1).

$$\Delta x = v.t = c_x.t$$

So, the beam reflected on the beam-splitter must necessarily deflect in the direction of motion such that its speed component ($c_x$) in that direction matches that of the inertial system ($c_x = v$).

$$c_x = c.\sin \alpha \quad \text{thus:} \quad \sin \alpha = c_x / c = v / c$$

Actually, the physical law governing the deflection is rooted on the transfer of momentum, so:

$$\sin \alpha = p_x / p_o = mc_x / mc = c_x / c = v / c$$

where $p_o$ is the momentum of the incident photons and $p_x$ their remnant component in the direction of the inertial system motion, after reflection on the beam-splitter.

### *2. Second order equivalency (or interferometric equivalency)*

### *2.a. Round-trip time-of-flight of the transmitted beam along the X axis* (fig.2)

Actual length along the X axis run by the beam in its round-trip from and back to the beam-splitter:

$$x_1 = c.t_1 = d + v.t_1 \qquad \text{and} \qquad x_2 = c.t_2 = d + v.t_2$$

Therefore the corresponding round-trip time-of-flight is:

$$t_x = t_1 + t_2 = d / (c - v) + d / (c + v)$$

Let us remark that the speeds c and v subtract and add algebraically. This formulation can be rewritten in the form:

$$t_x = t_1 + t_2 = [(d / c) / (1 - v / c)] + [(d / c) / (1 + v / c)] = 2.(d / c)/(1 - v^2/c^2)$$

*2.b. Round-trip time-of-flight of the reflected beam along the Y axis* (Fig.1)

Effective length run by the beam reflected on the beam-splitter up to the top mirror:

$$L_y^2 = (c.t)^2 = d^2 + (v.t)^2 \qquad \text{so, for one-way:} \qquad t^2 = d^2 / (c^2 - v^2)$$

The round-trip time-of-flight is thus:

$$t_y = 2.(d / c) / \text{sqr}(1 - v^2/c^2)$$

*2.c. Inequality of the reflected and transmitted beams time-of-flight*

$$t_x = 2.(d / c).(1 + v^2/c^2) \qquad \text{and} \qquad t_Y = 2.(d / c).(1 + 1/2\ v^2/c^2)$$

and therefore the temporal mismatch is equal to: $\Delta t = t_x - t_y = (d / c).(v^2/c^2)$.

So, from a classical analysis based on ray optics the two beams impinge back on the beam-splitter at the same position but at a different time, there time-of-flight differing increasingly with speed. However, second order equivalency of inertial systems implies the two beams to recombine preserving the phase against any speed variation, thus requiring the two time-of-flight to remain constant. Special relativity has solved this apparent paradox appealing to the Lorentz transformation, directly applied to space and time. Let us now see if it can be solved within a transgressing framework. A priori such a framework based on discordant postulates should not provide the right answer, neither conform with the predictions from special relativity.

Let us stress that what actually measures Michelson's interferometer is the phase and not directly the time-of-flight. Thus, let us centre on how the device perceives the period and wavelength. Let us be aware that they constitute the actual units of time and length, settling through the phase the differential measure of the beams time-of-flight and path length. Hence it is crucial to know how these units are perceived on the inertial system, and for any interferometric analysis the deduction of the beams path length and their time-of-flight should imperatively be made using these mutable units and not the standard ones which are fix units. When these units are used the beam time-of-flight and path length stay invariant under any variation of the system speed.

*2.d. Period and wavelength perceived by the inertial system* (fig.1 and 2)

In the direction of motion (X axis) the periods perceived by the system of the beam forward and backward paths AB and BA' (fig.2) are:

$$T_{x1} = \lambda / (c - v) \qquad \text{and} \qquad T_{x2} = \lambda / (c + v)$$

Since the intrinsic speed of light through space is equal to c, the respective wavelengths are:

$$\lambda_{x1} = c.T_{x1} = c.\ \lambda / (c - v) = \lambda / (1 - v / c) \qquad \text{and} \quad \lambda_{x2} = c.T_{x2} = c.\ \lambda / (c + v) = \lambda / (1 + v / c)$$

To account for the null interferometric result of Michelson's experiment, i.e. the independence from speed of the fringe pattern, one must logically count the number of period ($N_x$) perceived by the system during the beam round-trip time-of-flight.

$$N_{x1} = t_{x1} / T_{x1} = [d / (c - v)] / [\lambda / (c - v)] = d / \lambda$$

$$N_{x2} = t_{x2} / T_{x2} = [d / (c + v)] / [\lambda / (c + v)] = d / \lambda$$

$$N_x = N_{x1} + N_{x2} = 2.(d / \lambda)$$

In the direction perpendicular to the system motion (Y axis) the period and wavelength perceived by the inertial system of the beam forward path AB and backward path BA' (fig.1) are:

$$T_{y1} = (\lambda / c) / sqr(1 - v^2/c^2) \quad \text{and} \quad \lambda_{y1} = c. T_{y1} = \lambda / sqr(1 - v^2/c^2)$$

Since the beam round trip time-of-flight $t_y = 2. t_{y1} = 2.(d / c) / sqr(1 - v^2/c^2)$, the resultant number of periods perceived by the inertial system is:

$$N_y = t_y / T_{y1} = 2.[(d / c) / sqr(1 - v^2/c^2)] / [(\lambda / c) / sqr(1 - v^2/c^2)] = 2.(d / \lambda)$$

It appears that the number of periods perceived by the inertial system along the X and Y axis is exactly the same: $N_x = N_y = 2.(d / \lambda)$ and thus the two split beams preserve the same phase upon recombination, independently of the system speed.

Accounting for both, the beam time-of-flight and the actual unit of time embodied in the wave's period, as perceived by the system, it comes out that their number is independent of the system speed and always determined by the ratio $(d / \lambda)$. So, when the speed of an inertial system changes, the time of flight and the period evolve identically and so the corresponding number of periods stays invariant and hence the phase. Let us point out that the current approach stands in taking into account a well established speed-dependent physical effect, namely the Doppler effect, instead of a mathematical handling of time and space by means of the Lorentz transformation. So, from the transgressing framework the wavelength and period are the basic units of length and time, and these are the physical entities which vary, and not space and time themselves.

*2.e. Interferometric invariance*

The expected phase shift between the two beams upon recombination was:

$$\Delta L = c.(t_x - t_y) = c.(d / c).(v^2/c^2) = d.(v^2/c^2)$$

However, this expectation presumes the equality of the units of time along the X and Y axis, i.e. of the period $T_x$ and $T_y$ perceived by the inertial system. But this assumption is quixotic since the proper conveying system is actually moving through space, not only the beam. So, when the speed dependence of the period is accounted for, it appears that the number of periods of the transmitted and reflected beams are equal ($N_x = N_y$) and stays invariant whatever the system speed. What emerges thus from this analysis is that Michelson's interferometer being blind to speed, is unsuitable for its pretended purpose.

So, the conventional extrapolation that the speed of light is constant, independently of the observer speed lacks of analytic rigor, besides digressing rationality. Contrarily to the expectations put on Michelson's interferometer, these do not stand an accurate analysis that must include the Doppler effect, since it just cannot be arbitrarily omitted that the proper system is actually moving. The only conclusion wisdom upholds is that the interferometric pattern is independent of the system speed, due to the phase invariance resulting from the constancy of the number of periods, as a result of their speed dependence. So, it is not space and time which actually vary, as assumed by special relativity, but the units of space and time embodied in the wavelength and period. This confusion has lead to a grievous misleading conception of reality, allowed by the fact that mathematics do not catch the difference. The transposition of mathematical tools, such as most particularly space contraction, to an actual physical reality should be carefully questioned in view of the risk of inadvertently being building up a pseudo-reality. Besides, whatever the preferred conception, the conceptual dilemma arising from two mutually exclusive standpoints to provide the same result should be scrutinised.

*Conclusion*

Let us stress that the transgressing postulates provide the same result as that of special relativity. The emerging dilemma stands on the two sets of postulates being incompatible. So, it appears imperative to find an explanation to this incongruence. While special relativity applies a dilatation state to time and a contraction state to space itself, *the transgressing framework applies contraction and dilatation effects to the units of space and time, respectively physically embodied in the beam wavelength and period*. Mathematical constructs accept indifferently both formulations and provide identical results, so their valuation must be made on physical grounds. The point is that these mathematically equivalent layouts, since providing matching outcomes, lead however to completely different conceptions of physical reality: one leads to a concurrent multiplicity of states of space, while the other one appeals to the relative perception of the wavelength and the period on inertial systems. Let us also mention that, on physical grounds, it is quite painful to conceive how space could have coexistent distinct states from different systems speed.

The mathematical efficiency of special relativity is not at all brought into play (**) in spite of its confusion upon the actual cause of the equivalency of inertial systems, however our underlying concern is the safeguard of physical realism by taking care of avoiding the construct of imaginary worlds with fictitious features, based on a pseudo-reality aberrantly derived from mathematical frameworks. For example, let us somewhat analyse the concept of space contraction and its implications. What should be understood by the generic contraction of space, that of the whole universe? This is what it implies as long as a limited range of contraction is not justified, and thus the whole universe is concurrently in multiple different states of contraction, one for each inertial body, from elementary particles to cosmic bodies. So, special relativity accounts for the equivalency of inertial systems at the cost of leading to the inequivalency of space, i.e. a multiplicity of concurrent contraction states of space, one for each inertial system. Such a conceptual construct may bring a doubtful intellectual gain. Is it clever to extract space contraction from its mathematical framework and to extend its conception beyond being a mathematical wild card? Shouldn't be wiser not to transcript it to an actual physical reality?(***).

Furthermore, special relativity postulates the constancy of the speed of light relative to any inertial system. But, Michelson interferometer as well as its modern versions as optical cavities are indeed not evidencing the invariance of the relative speed of light as commonly thought, but are instead highlighting the incapacity of this type of optical devices to detect any net effect of speed, therefore being inadequate for their presumed objective. Improvement of their accuracy just better sound their degree of blindness to the conveying system speed. Extrapolating that the speed of light is constant relative to any inertial observer lacks of deductive rigor besides being irrational. It can only be deduced that such optical systems are unable to measure the speed of light relative to the inertial observer and instead *provide invariably the intrinsic speed of light through space*, independently of the conveying system motion. This confusion has been very harmful on conceptual grounds, by introducing pseudo-effects such as the contraction of space (13), which in fact is circumscribed to the mathematical construct which handles it. Here it has been shown that there is no need of appealing to these effects to give account of the electromagnetic equivalency, and space is indeed formally independent of the speed of inertial systems. It is essential to focus the analysis on the relationship between light and the optical set-up, i.e. in the properties of light as perceived by the observer and not to extrapolate these properties of light to properties of space and time.

The constancy of the speed of light relative to the observer and the consequent concept of contraction of space may be used as mathematical master keys, but are however considered to falsify physical reality. In effect, the electromagnetic first and second order invariance of inertial systems has been straightforwardly accounted for through the standard addition of speeds, once the Doppler effect is taken into account. There is no need to dismiss the ordinary logic since *it is not the speed of light relative to any inertial observer that is constant, but actually it is the phase that is oblivious to speed.* We have expressed and justified our opinion and our conceptual criterion which is the safeguard of rationality and realism, in contrast to the conceptual surrealism of special relativity. Now it is the reader's turn to ponder on interpreting physical reality through the contraction of space and dilatation of time or through the variation of the units of space and time, embodied in the wavelength and the period. These two issues, mathematically equivalent, lead however to drastically different conceptions of our physical world.

*Comments*

(\*) Special relativity applies to other ambits such as to clocks. Indeed, pions or muons, seen as elementary clocks, do exhibit a dilatation of their lifetime with speed. From a mathematical approach this effect has been attributed to the dilatation of time itself, however from a physical approach this effect is attributed to a decrease of the probability of disintegration, i.e. of the system's transition frequency from one state to another one. This equally applies to atomic and nuclear clocks. So, in all cases what is actually measured is the rate of transition as a function of speed, which effectively slow down. From it, the transition period is deducted, similarly to the electromagnetic case in which it is the period of the wave that is defined as a function of speed. So, the affectation of the period by speed applies in all cases, and in all measures of time the actual unit is the period which is a speed-dependant unit.

(\*\*) The mathematical efficiency of special relativity, which is not questioned, is based on the fact that it does not specify to which specific physical entity it should be applied, relying instead on generic parameters represented by space and time. Certainly this offers great advantages on mathematical grounds, since there is then no need to know the specificity of each experiment, but in counterpart this has lead to the naive belief in the contraction of empty space (13), which is today's "blind point" and is rooted on the endemic confusion between map and territory.

(\*\*\*) Within the framework of special relativity, when considering the case of atmospheric muons, for them, space has shortened due to their high speed, while for any external observer it is instead time that has slow down. Consequently, two different causes are attributed to the same phenomenon. This is an evident example of causal schizophrenia in which two different causes are attributed to the same effect, depending on the proper or outer observer, and which is however dogmatically defended on theoretical grounds and tought in Universities (13). From the author's point of view, the belief in the contraction of space as an actual physical effect is as erred as the belief of medieval scolastics in the epicentrality of earth by relying on Aristotle's doctrine and the Ptolemaic formalism.

*Author's other related articles:*

A dual set-up based on Bradley's aberration of light, using simultaneously stellar and local light sources.

A causal approach to first-order optical equivalency of inertial systems, by means of a beam-pointing test-experiment based on speed-induced deflection of light.

lanl.arXiv.org e-Print archive (url: xxx.lanl.gov), Physics, Subj-class: General Physics